# Onset Properties of Supersolid Helium

Yongle Yu
*State Key Laboratory of Magnetic Resonance and Atomic and Molecular Physics,
Wuhan Institute of Physics and Mathematics, Chinese Academy of Sciences, Wuhan 430071, P. R. China*
(Dated: April 30, 2010)

Supersolid helium has a rather low transition temperature and a small critical velocity, compared with liquid helium. These properties could be explained in terms of helium's spectrum structure and quantum jumps involving large momentum transfer. A grain in the solid helium possess valleys (local minima) in its many-body dispersion curve, and an exchange of large momenta with the grain's surroundings occurs in a jump between a level in one valley and another level in the neighboring valley. Such jump process also naturally causes dissipation accompanying the onset of supersolidity.

Quantum mechanics formulates that a system has its own eigen-energy levels, and that physical processes correspond to jumps among these levels. These fundamental formulations are essential to our understanding of various quantum phenomena, for example, atomic spectroscopy, and the difference between metals and insulators. In the same spirit, it can be shown that the spectrum of a superfluid is crucial to explain superfluidity and its related properties (see e.g. [1–6]). In this note, we argue that some quantum jumps involving large momentum transfer are responsible for the relative small critical velocities and low transition temperature in supersolid helium experiments [7–19]. We also explain the energy dissipation behavior of supersolid helium [14, 15, 17, 18] in terms of these jumps.

Let us consider a superfluid composed of $N$ bosons, which has a weak coupling with its surrounding environment. The total Hamiltonian of the superfluid and its environment can be written in the form of

$$\hat{H} = \hat{H}_{sf} + \hat{H}_{env} + \lambda \hat{H}_{coupling} \qquad (1)$$

where $\hat{H}_{sf}$, $\hat{H}_{env}$ are the Hamiltonian of the superfluid and that of the environment, respectively, and $\lambda \hat{H}_{coupling}$ is the coupling between them, with $\lambda$ being small to describe the weakness of the coupling.

For simplicity, we assume that the superfluid is translational invariant and it has periodic boundary conditions. $\hat{H}_{sf}$ has the form of,

$$\hat{H}_{sf} = -\frac{\hbar^2}{2M} \sum_i \frac{\partial^2}{\partial x_i^2} + \sum_{i<j} V(x_i, x_j) \qquad (2)$$

Where $M$ is the mass of a boson, $V$ is the two-boson interaction and $x_i$ is the coordinate of $i$th boson.

We denote $\psi_n(x_1, x_2, ..., x_N)$ the eigen wave function of $\hat{H}_{sf}$ where $n$ is the labeling index (equivalent to a set of $N$ quantum numbers). For the system in state $|\psi_n\rangle$, the corresponding eigen energy is given by $E_n = \langle\psi_n|\hat{H}_{sf}|\psi_n\rangle$ and the momentum by $P_n = \langle\psi_n|\hat{P}|\psi_n\rangle$, where $\hat{P} = \sum_i \hbar \frac{\partial}{\partial x_i}$ is the total momentum operator of the superfluid.

The state of the superfluid is described by the occupation probability of the eigen levels, $\rho_n(t)$, as functions of the time $t$. One has $\sum_n \rho_n(t) = 1$. The system has energy $E(t) = \sum_n \rho_n(t) E_n$ and momentum $P(t) = \sum_n \rho_n(t) P_n$ at the time $t$.

In general, changes of occupation probability is governed by the following equation,

$$\frac{d\rho_n}{dt} = \sum_{n' \neq n} A_{n' \to n} \rho_{n'} - \sum_{n' \neq n} A_{n \to n'} \rho_n \qquad (3)$$

Where $A_{n' \to n}$ is the transition rate of the superfluid from $|\psi_n\rangle$ to $|\psi_{n'}\rangle$, resulted from the coupling between the superfluid and its environment, and depending on the states of its environment.

Some rough analysis of $A_{n' \to n}$ can be done using standard perturbation methods. First, $A_{n' \to n}$ is negligible if $|\psi_n\rangle$ differs significantly from $|\psi_{n'}\rangle$ (The meaning of difference will be specified). A general form of coupling between the superfluid and its surroundings can be taken as $\lambda H_{coupling} = \lambda \sum_p f(p) \sum_k c^\dagger_{p+k} c_k \hat{D}_{-p} + c.c.$ where $c^\dagger_k (c_k)$ is the creating (annihilating) operator of a boson with momentum $k$, $\hat{D}_{-p}$ operates on the states of the environment and deposits a momentum of $-p$ into it, and $\lambda f(p)$ is the strength of a momentum exchange by $p$ between them. In first order perturbation, $A_{n' \to n} \propto \lambda^2 |\langle\psi_n| \sum_k c^\dagger_{q+k} c_k |\psi_{n'}\rangle|^2$ ($q$ is the momentum difference between $|\psi_n\rangle$ and $|\psi_{n'}\rangle$), thus $A_{n' \to n}$ vanishes unless $|\psi_n\rangle$ has significant overlap with $\sum_k c^\dagger_{q+k} c_k |\psi_{n'}\rangle$. The state $c^\dagger_{q+k} c_k |\psi_{n'}\rangle$ differs $|\psi_{n'}\rangle$ by an particle hole excitation. If $c^\dagger_{q+k} c_k |\psi_{n'}\rangle$ approximates another many-body eigenstate of $H_{sf}$, we might call it to be a particle-hole state with respect of $|\psi_{n'}\rangle$. Thus, in the first order perturbation, $|\psi_n\rangle$ should be particle-hole-like with respect to $|\psi_{n'}\rangle$ for that $A_{n' \to n}$ is non-negligible.

The higher order contributions of $A_{n' \to n}$, might being significant, can be analyzed similarly. For example, the second order contribution is non-negligible only when $|\psi_n\rangle$ has large overlap with $c^\dagger_{q-p_1+k_2} c_{k_2} c^\dagger_{p_1+k_1} c_{k_1} |\psi'_n\rangle$ [20] (for some $p_1, k_2, k_1$) which can be loosely called the two-particle-hole state with respect to $|\psi_{n'}\rangle$. The third order contribution is almost zero unless $|\psi_n\rangle$ can be approached by three-particle-hole excitation of $|\psi_{n'}\rangle$. The significant of $k$-th order contribution of $A_{n'\to n}$ requires that $|\psi_n\rangle$ can be approached by $k$-quasiparticle-hole excitation of $|\psi_{n'}\rangle$. The (minimum) numbers of particle-hole excitation for bridging between $|\psi_{n'}\rangle$ and $|\psi_{n'}\rangle$ can be referred as their degree of difference.

$A_{n'\to n}$, involving the parallel (many-body) state transitions of the environment, has a factor being a sum of probability of possible transitions in the environment with proper weight. The probability of the state transition in environment can be analyzed similarly as above. The transition is mediated by the corresponding particle-hole excitation of the many-body (eigen) states of the environment in the first order process, and higher order transition requires the final state and the initial state can be bridged by the corresponding higher particle-hole excitations.

A third important aspect of $A_{n'\to n}$ is the requirement of momentum-energy conservation for the quantum exchange between the superfluid and its surroundings. For exchange process involving small momentum exchange (momentum is conserved), the possible energy mismatch in the process could be small, for example, being within the range determined by (energy-time) uncertainty principle. However, for a process involving large momentum exchange, the energy mismatch is generally large, therefore this kind of process is likely prohibited.

The many-body dispersion spectrum of the superfluid is not monotonic [2–4] (see e.g. the dispersion curve in Fig. 1), and this feature causes different low temperature transport behaviors from those of a normal system. Fig. 1 schematically plots the spectrum of the superfluid at strong interaction (, i.e., a supersolid) [5], where the $x-$,$y-$ coordinate of a level represent its moment and energy, respectively. At sufficient low temperature, only the levels close to the bottoms of the valleys (located at the local minima of the curve) are important to determine the transport properties. It is naturally to consider the quantum jumps between neighboring valleys, which involves a change of large momenta $\sim \frac{\hbar}{a}$ where $a$ is the average neighboring distance of bosons in the system [5]. Depending on the transition rates of such inter-valley jumps, the phase transition properties are determined in different way: *case I*) inter-valley transition rate is negligible, phase transition is caused by thermal excitations which overcome the energy barriers between the valleys [6]; liquid helium belongs to this case; *case II*) The rate is non-negligible, then the transition temperature and critical velocity, determined in a way differently from *case I*.

In *case II*, understanding of transitional temperature and critical velocity of superfluidity requires level-specific analysis of the interval-valley jump. First, the occupation probability of levels at the bottom of a valley is roughly determined by the thermal equilibrium ,i.e., $\rho(E_n) \propto e^{-E_n/kT}$. An (effectively) occupied level, labeled by $n_0$ and with its energy denoted by $E_{n_0}$ and momentum by $P_{n_0}$, in the processes of large quantum jumps, may be connected to a group of levels at the next valley indexed by $m$, $m = 1, 2, ...$, i.e., the transition rate $A_{n_0\to m}$ is non-negligible. With the momentum and energy of level $m$ are denoted by $P_m$, $E_m$, respectively, the critical velocity for level $n_0$ is the minimum of $(E_m - E_{n_0})/(P_m - P_{n_0})$ among all $m$ (see Fig. 2). The critical velocity for a valley is then roughly the average of the critical velocities of the effectively occupied levels in the valley. Generally the levels with high energies have small critical velocities, and this make the critical velocity to be temperature dependent. Overall, in *case II*, both critical velocity and transition temperature could be much small compared with *case I* where the large momentum exchange process is absent.

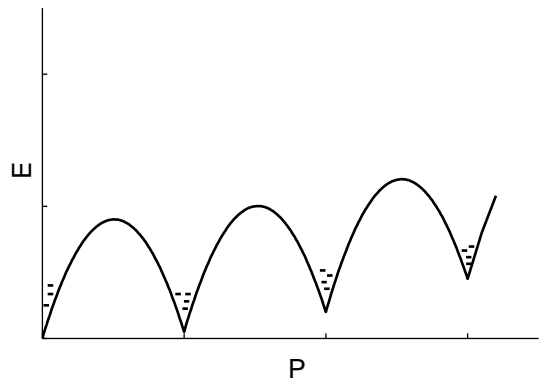

FIG. 1: A schematic plot of non-monotonic dispersion spectrum (line) with valley-like local minima for a superfluid. some excited states near the local minima of dispersion are marked by the bars. The low lying states close to bottoms of the valleys determine the low temperature transport behaviors.

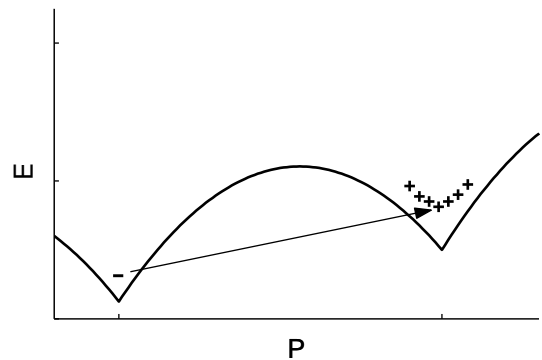

FIG. 2: A supersolid may jump from a low lying level (bar) at a valley to a group of levels (pluses) in the next valley, the jump marked by the arrow determines the critical velocity of the low lying level. The spectrum refers to a lab system.

The experimentally observed supersolid helium corresponds to *case II*. The helium system is generally confined within an annular regime and generally is composed of grains. Some grains in touch with container could move under the motion (rotation) of the container. Generally the force to a touching grain, exerted by the 'rotating' container, sometimes is not parallel to the grain's facet in touch with the container, and the grain could move under the partial normal force applied to the facet [21]. Inside the solid helium, grains interact with their neighboring grains. Some grains, which may not move with the same velocity with their neighbors, are the superfluid fraction of solid. It is then 'exceptionally' possible that these supersolidic grains can exchange large momentum with their neighbors (in the normal phase). The 'exception' is due to that these grains have similar many-body level structure, and thus energy mismatch at large momentum transfer is not an issue.

The experimentally observed dissipation of torsional oscillator, accompanying the onset of supersolidity, can be accounted by the behavior of internal energy of the supersolidic grains. In the normal phase, the inter-valley jump is efficient enough so a supersolidic grain jumps fast from the initial occupied valley(s) to the valley(s) where it has the same velocity as its neighbors. With the occupation probabilities of levels in the final valley(s) being thermalized, the energy of the grain, apart from its kinetic energy, doesn't change much. In the supersolid phase, the inter-valley jump is negligible and a supersolidic grain just stays in the initial valley and its internal energy remains the same. Thus away from the onset regime, there is not much energy dissipation of the grain. At the onset stage of supersolidity, a grain still jumps from valleys to valleys. However, due to the small transfer rate, the partially occupied states are distributed in many valleys. Such a level distribution corresponds to a nonequilibrium state and the internal energy of the grain is increasing, thus it absorbs energy from its surroundings and causes the dissipation of the torsional oscillator. In literature, glass model [22], and superglass model [23], are proposed to explain the dissipation behavior. One can naturally realize that glass and supersolid share a common feature in their level structures: the possession of many metastable states.

In conclusion, various onset properties of supersolid helium can be microscopically understood in terms of quantum jumps among the levels of the system.

[1] L. D. Landau, J. Phys. USSR **5**, 71 (1941).
[2] F. Bloch, Phys. Rev. A **7**, 2187 (1973).
[3] A. J. Leggett, Rev. Mod. Phys. **73**, 307 (2001).
[4] Y. Yu, Ann. Phys. **323**, 2367 (2009).
[5] Y. Yu, cond-mat/0609712v2.
[6] Y. Yu, arXiv:0908.1002.
[7] E. Kim and M. H. W. Chan, Nature **427**, 225 (2004).
[8] E. Kim and M. H. W. Chan, Science **305**, 1941 (2004).
[9] E. Kim and M. H. W. Chan, Phys. Rev. Lett. **97**, 115302 (2006);
[10] A. C. Clark, J. T. West, and M. H. W. Chan, Phys. Rev. Lett. **99**, 135302 (2007);
[11] A. Penzev, Y. Yasuta, and M. Kubota, J. Low Temp. Phys. **148**, 667 (2007);
[12] M. Kondo, S. Takada, Y. Shibayama, and K. Shirahama, J. Low Temp. Phys. **148**, 695 (2007).
[13] Y. Aoki, J. C. Graves, and H. Kojima, Phys. Rev. Lett. **99**, 015301 (2007).
[14] A. S. C. Rittner and J. D. Reppy, Phys. Rev. Lett. **97**, 165301 (2006).
[15] A. C. Rittner and J. D. Reppy, Phys. Rev. Lett. **98**, 175302 (2007).
[16] M. W. Ray and R. B. Hallock, Phys. Rev. Lett. **100**, 235301 (2008).
[17] A. Penzev, Y. Yasuta, and M. Kubota, phys. Rev. Lett. **101**, 065301 (2008).
[18] J. T. West, X. Lin, Z. G. Cheng, and M. H. W. Chan, phys. Rev. Lett. **102**, 185302 (2009).
[19] M. W. Ray and R. B. Hallock, J. Low. temp. Phys. **158**, 560 (2010).
[20] We here ignore the operator nature of the factor $\frac{1}{E-H}$ in the normal perturbation expansion form $...\frac{1}{E-H}H_{coupling}\frac{1}{E-H}H_{coupling}...$.
[21] Y. Yu, arXiv:0907.3971.
[22] Z. Nussinov, A. V. Balatsky, M. J. Graf, and S. A. Trugman, Phys. Rev. B **76**, 014530 (2007).
[23] B. Hunt, E. Pratt, V. Gadagkar, M. Yamashita, A. V. Balatsky, and J. C. Davis, Science **324**, 632 (2009).